\documentclass[12pt]{article}%
\usepackage{amssymb}
\usepackage{amsmath}
\usepackage{amsfonts}
\usepackage{sw20ssur}
\usepackage{graphicx}
\usepackage{accents}%
\providecommand{\U}[1]{\protect\rule{.1in}{.1in}}
\newtheorem{theorem}{Theorem}

\newtheorem{proposition}[theorem]{Proposition}

\ifx\pdfoutput\relax\let\pdfoutput=\undefined\fi
\newcount\msipdfoutput
\ifx\pdfoutput\undefined\else
\ifcase\pdfoutput\else
\ifx\paperwidth\undefined\else
\ifdim\paperheight=0pt\relax\else\pdfpageheight\paperheight\fi
\ifdim\paperwidth=0pt\relax\else\pdfpagewidth\paperwidth\fi
\fi\fi\fi
\begin{document}

\title{Is cosmic dynamics self-regulating?\\\ \ \ }
\author{Manasse R. Mbonye\thanks{\ ICTP-East African Institute for Fundamental
Research, University of Rwanda, Kigali, Rwanda} \thanks{\ Contact:
mmbonye@gmail.com; mmbonye@eaifr.org}\ \ \\\ }
\maketitle

\begin{abstract}
In this paper we discuss a cosmological model for a universe with
self-regulating features. We set up the theoretical framework for the model
and determine the time evolution of the scale-factor $a(t)$. It is shown that
such a universe repeatedly goes through alternate periods of matter and dark
energy domination. The resulting dynamics oscillates about the would-be ideal
time-linear or coasting path, with monotonic expansion. When compared to
dynamics of the observed physical Universe, the model recovers the
observationally-established evolutionary features of the latter, from the big
bang to the current acceleration, and farther. It suggests a universe that
initially emerges from a non-singular state, associated with a
non-inflationary acceleration, and which acceleration it exits naturally with
matter-energy generation. The model does not have a horizon problem or a
flatness problem. It reproduces the observed current values of standard cosmic
parameters, including the age $t_{0}$, the current Hubble parameter $H_{0}$
and dark energy $\Omega_{de}\ $and matter $\Omega_{m}$ density parameters. We
find the dark matter density-profile generated by the model naturally leads to
flat rotation curves in galaxy halos. The model is falsifiable. It makes
predictions that can be tested, as suggested. Finally, we discuss the
dimensionless age $(H_{0}t_{0}\simeq1)$ paradox as an example of the model%
\'{}%
s ability to address standing puzzles. The findings suggest dynamics of the
physical Universe may be self-regulating and predictable.

\end{abstract}

\section{Introduction}

A little over two decades ago observations indicated [1][2] that the Universe
has been in a state of acceleration for the last four billion years or so [3].
This acceleration is explained to result from dark energy, a field with
negative pressure. According to $\Lambda CDM$ [4], the current leading model,
this field has a constant energy density. The $\Lambda CDM$ model, which has
its roots in the inflation paradigm [5], has been quite successful in agreeing
with more observations than most other approaches [6]. Still there are
challenges and unexplained issues. They include, an unexplained emergence of
dark energy in cosmology [7], a need for a clear link between the current
accelerated phase and previous cosmic phases, and a need for predicting the
future of cosmic evolution, in general. There are also observations that are
still unexplained. They include the recent James Web Telescope (JWST)
observation of galaxies at earlier than expected cosmic ages \textbf{[8]},
along with evidence of small and compact star distribution at low redshifts
\textbf{[9]}. Then there are several observational puzzles, such as the
synchronicity or dimensionless age problem \textbf{[10]}. These issues suggest
either a need for significant improvement of the existing model or a need for
a paradigm shift. Several new models based on GR have lately been suggested,
including for example [11], [12], [13]. For a an exhaustive review (see [14]).

In the present work we study dynamics of a universe with self-regulating
features. The model which is based on General Relativity (GR) \textbf{[15]}
explores the consequences of introducing a couple of ingredients into the
existing Friedman-Lamaitre-Robertson-Walker (FLRW) model (see \textbf{[16]}).
The first ingredient is based on the accepted concept (in GR) that empty
physical space deforms under the influence of matter-energy. Our
interpretation is that if empty space (hereafter\textit{ basic-space}) can
deform it must have a structure. In this model we show how based on this
concept of its structure basic-space emerges as the main driver for the
dynamics of the modeled universe. The second ingredient is that free
matter-energy emerges as a perturbation to basic-space, and hence constituting
a perturbation to the latter`s dynamics. Thus, in the model, matter-energy
tends to shift basic-space from its preferred dynamics, which we show to
constitute a coasting path. It is shown that in a bid to remain stable to such
perturbations, and to recover its dynamics, basic-space reacts by setting up a
restoring tendency by creating matter of a gravitational signature opposing
that of the perturbations. We set up the theoretical framework for the model
and determine the time evolution of the scale-factor $a(t)$. In the resulting
dynamics the system oscillates about the would-be ideal time-linear or
coasting path, with monotonic expansion. We compare the results of the model
with the features and dynamics of the observed physical Universe and discuss
the results of the comparison.

The rest of the paper is arranged as follows. Section 2 discusses basic-space
and its contributions to cosmic dynamics in the model. Section 3 discusses
contributions to cosmic dynamics from free matter-energy of the model. Section
4 lays out the theoretical and observational basis for the approach. Section 5
combines the two ingredients to construct a full cosmic dynamics of the model.
Section 6 establishes the model's consistency with observations of the
physical universe. Section 7 concludes the paper.

\section{A model of basic-space and its contributions to cosmic dynamics}

In this section we construct the dynamics of basic-space for the model based
on Friedman cosmology which we first summarize below.

\subsection{Friedman cosmology}

In Friedman-Lamaitre-Robertson-Walker (FLRW) cosmology cosmic dynamics is
driven by matter-energy which, at large scale, constitutes a homogeneous fluid
with isotropic expansion. This constitutes the statement of the Cosmological
Principle. The geometry is described in commoving coordinates by the
Robertson-Walker metric [16] with a line element,%

\begin{equation}
ds^{2}=dt^{2}-a\left(  t\right)  ^{2}\left(  \frac{dr^{2}}{1-kr^{2}}%
+r^{2}d\Omega^{2}\right)  , \tag{2.1}%
\end{equation}
where $a(t)$ is the cosmic scale-factor and $k=\left\{  +1,0,-1,\right\}  $
scales positive, flat or negative spatial curvature sections, respectively.
The FLRW metric $g_{\mu\nu}$, giving rise to the line element in Eq 2.1, is a
solution to the Einstein field equations $G_{\nu}^{\mu}=8\pi T_{\nu}^{\mu}$,
for a perfect fluid $T_{\nu}^{\mu}=diag(\rho,-p,-p,-p)$ of density $\rho$ and
pressure $p$. Here, the respective time and space components of the field
equations, respectively, read as%

\begin{equation}
\frac{\dot{a}^{2}}{a^{2}}+\frac{k}{a^{2}}=\frac{8\pi G}{3}\rho, \tag{2.2}%
\end{equation}
and
\begin{equation}
2\frac{\ddot{a}}{a}+\frac{\dot{a}^{2}}{a^{2}}+\frac{k}{a^{2}}=-8\pi Gp.
\tag{2.3}%
\end{equation}
From Eq. 2.2 and Eq. 2.3 one obtains
\begin{equation}
\frac{\ddot{a}}{a}+\frac{4\pi G}{3}\left(  \rho+3p\right)  =0. \tag{2.4}%
\end{equation}
In cosmology, Eq. 2.2 and 2.4 constitute the Friedman equations. Taking time
derivatives of Eq. 2.2 and working with Eq. 2.4 also gives the (not
independent) conservation equation
\begin{equation}
\dot{\rho}+3\frac{\dot{a}}{a}\left(  \rho+p\right)  =0. \tag{2.5}%
\end{equation}

\subsection{basic-space}

As indicated by the time dependence of the scale-factor $a(t)$ in the
equations above, gravitational interactions of matter-energy with space, as
first predicted by Einstein [15] and observed by Hubble [17], lead to a
non-static and sometimes accelerating [1][2] universe. Indeed, whether it is
due to lumpy matter-energy fields as manifested by in gravitational lensing
[18], or by gravitational waves [19], free matter-energy interactions with
space lead to a spacetime that is either dynamic, curved, or both. This
observation that empty space, in response to matter-energy interactions, is
deformed by matter-energy logically suggests that basic-space does have a
physical structure. Such structure need be energy-like, akin to the free
energy fields that deform it. It also need be gravitationally neutral, with
net zero gravitational charge, which implies that basic-space in this model
need have net-zero spatial curvature, $k=0$. Our first challenge is to
quantify these two seemingly conflicting characteristics, as we do below.

The notion that space does have an underlying structure is not new. Such
concept is, for example, the basis for Loop Quantum Gravity [20][21]. Indeed,
the purpose of this paper is not to construct and/or discuss an elementary
theory of basic-space. Instead, here, we seek to identify expected emergent,
large scale, manifestations of the above-mentioned energy-like structural
characteristics of basic-space and whether (and how) they can influence the
dynamics of the universe in the model by making basic-space, itself, dynamic.

We first isolate these basic-space effects from those expected to be due to
free matter-energy fields (not part of basic-space). In consequent sections we
shall use this approach to take a new look at effects of free matter-energy
fields on cosmic dynamics, in the now dynamic basic-space background. Later,
we combine the two influences into a framework that leads to the full working
cosmic dynamics. To this end, we begin by supposing that at large
(macroscopic) length scales the emergent features due an energy-like structure
of basic-space (empty spacetime), effectively take on characteristics of a
perfect fluid of density $\sigma_{s},$ and pressure $\hat{\pi}_{s}$,
representable as a diagonal stress-energy tensor, $T_{\nu}^{\mu}%
=diag(\sigma_{s},-\hat{\pi}_{s},-\hat{\pi}_{s},-\hat{\pi}_{s})$. A fluid with
similar characteristics was previously utilized [23], in a different setting
by Kolb. The size scale of this effective fluid is identified by a
scale-factor, $a_{s}$. Ideally, in absence of free-matter energy fields to
dynamically deform it, the fluid rests in an inertial frame (which it
constitutes), so that $\ddot{a}_{s}=0$. Further, in satisfaction of the
gravitational neutrality as mentioned before we set $k=0$, which is also
consistent with persisting observations [22] that the physical Universe
appears virtually spatially flat. Applying these conditions to Eq. 2.2 and Eq.
2.3, respectively, we have that.%

\begin{equation}
\frac{\dot{a}_{s}^{2}}{a_{s}^{2}}=\frac{8\pi G}{3}\sigma_{s}, \tag{2.6}%
\end{equation}
and
\begin{equation}
\frac{\dot{a}_{s}^{2}}{a_{s}^{2}}=-8\pi G\hat{\pi}_{s}, \tag{2.7}%
\end{equation}

Eq. 2.6 and Eq. 2.7 constitute the working Einstein's field equations for
basic-space, here. As is evident above, the equations reflect a high symmetry.
This is a desired feature of basic-space. In particular, Eqs. 2.6 and 2.7
imply basic-space satisfies a barotropic equation of state $\rho=wp$ of the
form
\begin{equation}
\hat{\pi}_{s}=-\frac{1}{3}\sigma_{s}. \tag{2.8}%
\end{equation}
In turn, Eq. 2.8 implies the vanishing of the active gravitational mass
(charge) density, $\sigma_{s}+3\hat{\pi}_{s}=0$. This result both confirms and
guarantees the earlier condition of energy-neutrality of this spacetime, with
a non/vanishing energy density $\sigma_{s}$. Therefore basic-space can have a
structure that is both energy-like and gravitationally neutral, with no conflict.

\subsection{Dynamics induced by basic-space}

Integration of the conservation relation of Eq. 2.5 along with use of
$p=w\rho$ leads to $\sigma_{s}\propto a^{-3(w+1)}$ which (in our case) on use
of Eq. 2.8 implies $\sigma_{s}\ \sim\frac{1}{a_{s}^{2}}$. Comparison of this
result with Eq. 2.6 shows that the scale-factor grows linearly with time,
$a_{s}\sim t$ or%
\begin{equation}
a_{s}=\upsilon^{\ast}t, \tag{2.9}%
\end{equation}
where, for convenience, we denote inverse age as $\upsilon^{\ast}=\frac
{1}{t_{0}}$, so that $\dot{a}_{s}=\upsilon^{\ast}$ and $\left(  a_{s}\right)
_{0}=1$\textbf{. }We revisit this relation later (in Section 4) when setting
up boundary conditions for the full cosmic dynamics. Using Eq. 2.9 in Eq. 2.1
and $\ $for $k=0$ we get the familiar looking flat metric form
\begin{equation}
ds^{2}=dt^{2}-\left(  \upsilon^{\ast}t\right)  ^{2}\left(  dr^{2}+r^{2}%
d\Omega^{2}\right)  . \tag{2.10}%
\end{equation}
Eq. 2.10 is the formal solution to the Einstein equations of Eq. 2.6 and Eq.
2.7 for basic-space, in this model.

On cosmological scales the above result (Eqs. 2.10 and 2.9) depicts a constant
expansion-rate, flat spacetime, with a flat spatial section. As Eq. 2.9
indicates, basic-space will induce onto the Universe, time-linear or coasting
dynamics. Coasting cosmologies have previously been discussed, going back to
the Milne universe [24]. They include Kolb's K-matter universe [23], Melia's
$R=ct\ $universe [25], Benoit-Levy and Chardin's Dirac-Milne universe [7], and
the John and Joseph model [26], to name a few. For a recent review and
expanded list of contributions in the area, please see [27]. With exception of
Milne's curvature-driven universe, in all these treatments \textit{all the}
\textit{available} free matter-energy, in the relevant universe, is
incorporated in the active gravitational mass density $\rho+3p=0$ to produce a
coasting cosmology. On the other hand, in the present treatment the coasting
behavior is purely a characteristic of basic-space (or empty spacetime),
$\sigma_{s}+3\hat{\pi}_{s}=0$, $\sigma_{s}\neq0$, with neither curvature nor
free matter-energy fields playing any basic role. As noted below, the free
matter-energy contribution is considered separately, later. Further, our
present solution of basic-space, incorporating non-trivial elemental
energy-like structures at fundamental length scales, also differs from Milne's
solution both in content and geometry.\ We will take the result of Eqs. 2.10
and 2.9 as a general relativistic representation of this emergent basic-space,
which signifies an underlying elementary energy-like structure at fundamental
length scales. It is in this respect we believe that this result represents a
unique classical solution of basic-space with potential applications as
classical limit of elementary space considerations.

Finally, we point out that the discussed coasting behavior from basic-space is
only a background (and partial) contribution to cosmic dynamics, in this
treatment. As we show later, the full evolutionary dynamics of the universe in
the model will include contributions from free matter-energy fields\ and be
represented by a relation whose solution is a scale factor evolving as
\begin{equation}
a\left(  t\right)  =a_{s}+\delta a, \tag{2.11}%
\end{equation}
where $a_{s}$ is the contribution from basic-space and $\delta a$ is a
perturbative contribution from free matter-energy fields. In what follows, we
find $\delta a$, determine $a\left(  t\right)  $ and then work backwards to
find the equation for which $a\left(  t\right)  $ is a solution.

\section{Contributions to cosmic dynamics from matter-energy}

In this section a framework is set up to discuss the contribution of free
matter-energy to dynamics of the universe in this model. Later, we seek to
link such framework with the contribution to dynamics from basic-space
previously discussed, leading to a working general framework for the dynamics.

\subsection{Thermodynamics of matter-energy generation}

We start by considering a fluid of $N$ particles contained in some volume $V$.
From conservation requirements we have that $N_{;\mu}^{\ \mu}\equiv n_{;\mu
}u^{\mu}+n\Theta=0$. Here $n=\frac{N}{V}$ is the particle density, $\Theta$ is
the fluid expansion and $u^{\mu}$ is a 4-velocity of a commoving observer. If
the fluid source is matter generating, then
\begin{equation}
n_{;\mu}u^{\mu}+n\Theta=n\Gamma, \tag{3.1}%
\end{equation}
where $\Gamma$ is the creation rate. For positive gravitational charge,
$\Gamma>0$ implies matter-energy generation while $\Gamma<0$ implies
matter-energy annihilation [28][29].\ The concept of gravitationally-induced
cosmic matter generation was introduced in cosmology by H. Bondi [30] and F.
Hoyle [31] in their steady state model that later turned out unsuccessful.
Matter generation has since been applied in different approaches such as in
[32], [28] and [29]. Here, we start from the approach followed by [28] and [29].

For the above system of $N$ particles, with a density $\rho$ and pressure $p$,
the Gibbs entropy density evolves as%
\begin{equation}
Tds=d\left(  \frac{\rho}{n}\right)  +pd\left(  \frac{1}{n}\right)  . \tag{3.2}%
\end{equation}
Taking time derivatives of Eq. 3.2 and substituting for $\dot{n}$ using Eq.
3.1 we have $nT\dot{s}=\dot{\rho}+\Theta(\rho+p)-\left(  \rho+p\right)
\Gamma$. Further, if we assume quasi-adiabatic conditions, so $\dot{s}=0$,
then $\dot{\rho}+\Theta\rho+p-\left(  \rho+p\right)  \Gamma=0$. One can adopt
the process to cosmology by setting $\Theta=3H$, where $H=\frac{\dot{a}}{a}$
is the Hubble parameter. One then finds that $\dot{\rho}+3H(\rho+p+P_{c})=0$,
where $P_{c}=-\frac{1}{3H}\left(  \rho+p\right)  \Gamma$ is identified as the
creation pressure [28].

\subsection{Matter-energy generation by basic-space}

Specializing to the current approach of creation by basic-space, we have that%

\begin{equation}
\dot{\sigma}_{s}+3H_{m}(\sigma_{s}+\hat{\pi}_{s}+P_{c})=0, \tag{3.3}%
\end{equation}
where, as before, $\sigma_{s}$ and $\hat{\pi}_{s}$ are respectively the
density and pressure in Section 2, and where now%

\begin{equation}
P_{c}=-\frac{1}{3H_{m}}(\sigma_{s}+\hat{\pi}_{s})\Gamma\tag{3.4}%
\end{equation}
is the creation pressure of basic-space. Here $H_{m}=\frac{\dot{a}_{m}}{a_{m}%
}$, with $a_{m}$ identifying the influence on the evolution of the cosmic
scale-factor $a(t)$ by creation\footnote{Note that neither $H_{m}$ nor $a_{m}$
are independent physical quantities, on their own. Only $a(t)$ and $H(t)$ as
developed later in Section 4 are physical.}. Putting Eq. 3.4 back into Eq.
3.3, and applying Eq. 2.6 to the result we have that,%
\begin{equation}
\dot{H}_{m}+\frac{3}{2}\gamma H_{m}^{2}\left[  1-\frac{\Gamma}{3H_{m}}\right]
=0, \tag{3.5}%
\end{equation}
where, we have also used the general barometric pressure relation $\hat{\pi
}_{s}\left(  \sigma_{s}\right)  =\left(  \gamma-1\right)  \sigma_{s}$.
Comparison with Eq. 2.8 shows $\gamma=\frac{2}{3}$. Using this value in Eq.
3.5 gives $\dot{H}_{m}+H_{m}^{2}\left[  1-\frac{\Gamma}{3H_{m}}\right]  =0$,
and on substituting for $\dot{H}_{m}$ using $\ \dot{H}_{m}=\frac{\ddot{a}}%
{a}-H_{m}^{2}$ we get
\begin{equation}
\frac{\ddot{a}_{m}}{a_{m}}-\frac{1}{3}\Gamma H_{m}=0. \tag{3.6}%
\end{equation}
We make a couple of observations about this result. Since the Friedman
equation for basic-space (given by Eq. 2.6) here exclusively involves only one
term on the right, it is a definition and can be conveniently is used to
substitute for $\sigma_{s}$ and its derivatives in Eq. 3.3. This has two
effects. First it leads to Eq. 3.6 which is purely an evolution of the space
geometry. Secondly the background terms (of Eq. 2.6) self-cancel to leave Eq.
3.6 describing only the evolution of perturbations (of basic-space) terms. As
will be demonstrated shortly, it is in this sense that matter is introduced in
this model as a perturbation to background basic-space of Eqs. 2.6 and 2.7. In
the section that follows we will seek to solve Eq. 3.6, and later link the
results with those of unperturbed basic-space in Section 2 toward the model's
cosmic dynamics framework.

\section{Basis for regulated cosmic dynamics}

As the preceding brief discussion demonstrates, the dynamics associated with
gravitationally-induced matter-energy generation is set here as a perturbative
process on the background dynamics from the previous section, due to the
properties of basic-space. As we set to build the model's proposed cosmic
dynamics framework from these two ingredients, we first highlight on existing
observational and theoretical evidence that motivates and justifies the
approach to be taken.

\subsection{Observational and theoretical evidence}

Modern cosmology essentially rests on two pillars, namely: Einstein's General
Relativity whose field equations show the Universe can not be static [2], and
Hubble's observation that the Universe is, indeed, expanding [3]. Starting
from these pillars and evolving the Universe back in time led to the concept
of an initial cosmic state in form of a hot big bang [33]. One important
signature of the big bang was matter-energy created then as radiation,
predicted to currently form a cosmic microwave background (CMB) [34]. The
observation of CMB in 1962 by Wilson and Penzias [4] anchored the big bang as
a theory. The current theoretical and observational consensus, then, is that
matter-energy creation (in form of radiation soup) is associated with this
initial period. The original big bang theory left puzzling features that
included the flatness and horizon problems. The theoretical remedy by Guth [5]
was to introducing a primordial phase of cosmic inflationary acceleration that
preceded the big bang, driven by negative pressure of a constant net negative
gravitational mass density, $\rho_{\phi}+3p_{\phi}<0$, scalar field. The end
of this primordial accelerated phase is thereafter followed by a
(gravitationally positively charged) radiation, $\rho_{r}+3p_{r}>0$,
domination era, which eventually evolves into a cold matter, $\rho_{m}>0$,
dominated era that facilitates structure formation. Observations [1][2] now
show that this latter phase has, since about 4 billion years ago [3], given
way to the current cosmic acceleration, which according to theory [4] is
driven, by a dominating dark energy with negative pressure and with an overall
net negative gravitational mass density $\rho_{d}+3p_{d}<0$.

Thus, based on this observational and theoretical evidence it can be inferred
that (i) in the past the dynamics of the physical Universe has periodically
changed the sign of its acceleration, and that (ii) such change in the sign of
its acceleration was always associated with change in the sign of the
dominating gravitational mass density charge. While its first phases are
connected by reheat period [35] the current phase has neither any established
connection with past phases nor any known graceful exit to a future phases.

In what follows we present a simple framework to discuss cosmic dynamics,
grounded in the preceding observational and theoretical evidence. To proceed,
we begin with a set of propositions as a grounding for the framework, based on
evidence above of a universe with a history of alternating cosmic phases,
dominated and driven by matter-energy of correspondingly alternating net
gravitational mass density charge [36].

\subsection{The Dynamic Equilibrium Protection Proposal (DEPP)}

There are 3 attributes, based on preceding observational and theoretical
evidence, which we state below in form of propositions and thereafter apply to
link the\ model's two ingredients.

\begin{proposition}
An ideal universe constituted purely by basic-space remains in a state of
constant (coasting) expansion unless perturbed by free matter-energy fields.
\end{proposition}

\begin{proposition}
The universe creates stability against perturbations that tend to shift it
from its coasting, dynamic equilibrium state. (Such include density
perturbations growing from quantum fluctuations)\textit{ }
\end{proposition}

\begin{proposition}
\textit{When perturbed, the universe will suppress the perturbations through
creation of free matter-energy with a net gravitational charge opposite that
of the perturbations.}
\end{proposition}

\textbf{Explanation: }The attribute leading to \textit{Proposition 1} is a
consequence of the first ingredient, namely, that\textit{ }basic-space has
structure, whose effects are discussed in Section 2. It is essentially the
analogue of Newton `first law of motion. The attribute leading to
\textit{Proposition 3 }is, on the other hand, a consequence of the second
ingredient, and a consequence of the observational and theoretical evidence
just presented, that the universe can create matter-energy of either net
gravitational charge, whenever it suits it to. As we find later, it is also by
this proposition that, in search for dynamic stability, suppression of
perturbations through creation of opposite gravitational charges will proceed.
\textit{Proposition 2} establishes a linkage between attributes 1 and 2, and
hence a linkage between the two ingredients. As we shall show,
\textit{Proposition 2, }which is the analogue of Lenz` law of induction in
electrodynamics\textit{, }is also the central explanation to why the idealized
dynamic equilibrium state constitutes an attractor that ensures the observed
Universe is never too far from this ideal state. This could also a reasonable
starting point for a future resolution of the \textit{Coincidence Problem.}
The 3 propositions above constitute what is referred to, here, as the Dynamic
Equilibrium Protection Proposal, DEPP. We proceed to use DEPP to develop this dynamics.

\subsection{Regulation through annihilation}

Referring to the attribute implied in \textit{Proposition 3}, whenever the
universe\ in the model happens to be in a state dominated by a net positive
gravitational mass density $\rho_{r,m}+3p_{r}>0$, so that its dynamics is
characterized by cosmic deceleration,$\ \ddot{a}<0$, away from equilibrium,
then, in order to off-set such influence of net positive matter-energy
domination, and in search of restoring its dynamic equilibrium state, the
Universe will generate (or equivalently decay the former into) matter-energy
with net negative gravitational mass density $\rho_{d}+3p_{d}<0$, by
triggering $\Gamma<0$. Conversely, whenever the Universe is in a negative
matter-energy-dominated state $\rho_{d}+3p_{d}<0$ so that its dynamics is
characterized by cosmic acceleration, $\ddot{a}>0$, away from its equilibrium
state, then, in order to off-set this influence of negative matter-energy
domination, and in search of restoring its dynamic equilibrium state this
universe will generate (or equivalently decay the former into) net positive
matter-energy, $\rho_{r,m}+3p_{r}>0$, by triggering $\Gamma>0$.

In this sense, the Universe \textit{actually} utilizes the
gravitationally-induced creation rate, $\Gamma$, for purposes of
\textit{annihilation} of pre-existing matter-energy fields. This argument of
regulation by matter-energy annihilation and the above three propositions that
lead to it, form the basic argument for a regulated cosmic dynamics in the
model. This behavior is consistent with, and motivated by the observational
and theoretical evidence in the previous discussion.

Explicitly, the annihilation statement implies modification of Eq. 3.6 to now
take the form$\ \frac{\pm\ddot{a}_{m}}{a_{m}}-\frac{\mp\left\vert
\Gamma\right\vert H_{m}}{3}=0$. This can be re-written in a compact form as%

\begin{equation}
\frac{\ddot{a}_{m}}{a_{m}}+\frac{1}{3}\left\vert \Gamma\right\vert H_{m}=0,
\tag{4.1}%
\end{equation}
Eq. 4.1 sets the matter-energy regulatory controls on cosmic dynamics in the
model. In this treatment we adapt a condition on $\left\vert \Gamma\right\vert
$ such that $\left\vert \Gamma\right\vert \propto\frac{1}{H_{m}}$ (see
justification below). Related conditions have previously been discussed before
in the literature [28] [29], in different applications. In our case, the
choice is made by the realization $\left\vert \Gamma\right\vert $\ \ must tag
and hence couple to $H_{m}$ as implied in \textit{Proportion 3. }

Then, setting $\frac{1}{3}\left\vert \Gamma\right\vert H_{m}=\omega
^{2}=constant$, we can rewrite Eq. 4.1 in a familiar form of a\ classical
harmonic oscillator,
\begin{equation}
\ddot{a}_{m}+\omega^{2}a_{m}=0. \tag{4.2}%
\end{equation}
Eq. 4.2 describes the effects of the matter-energy perturbative contribution
to the time evolution of the cosmic scale-factor, in this model. This equation
admits harmonic solutions of the form:%

\begin{equation}
a_{m}=a_{\max}\sin\left(  \omega t+\psi\right)  , \tag{4.3}%
\end{equation}
where $a_{\max}$ is the maximum deviation (due to the perturbations), of the
scale-factor from its would-be equilibrium path, and $\psi$ measures the
initial $(t=0)$ phase angle of the implied cosmic oscillations. These 2
parameters will be constrained.

\subsection{Regulation by annihilation $\left\vert \Gamma\right\vert $, and
the cosmic period.}

In transforming Eq. 4.1 into relation of Eq. 4.2 we demanded that $\frac{1}%
{3}\left\vert \Gamma\right\vert H_{m}=\omega^{2}=constant$. This choice can be
independently reproduced. We set the creation /annihilation rate (absolute
value) to be proportional to the acceleration at a given expansion rate
$\left\vert \Gamma\right\vert \propto\frac{\ddot{a}_{m}}{\dot{a}_{m}}$,
consistent with propositions 3 and 2. Then $\left\vert \Gamma\right\vert
H_{m}=\kappa\left(  \frac{\ddot{a}_{m}}{\dot{a}_{m}}\right)  \left(
\frac{\dot{a}_{m}}{a_{m}}\right)  $ where $\kappa$ is a numerical
constant.\ Assuming harmonic solutions of the form $a_{m}=a_{\max}\sin\left(
2\pi\frac{t}{\tau}+\psi\right)  $, with a period $\tau=\frac{2\pi}{\omega}$,
we have that $\left\vert \Gamma\right\vert H_{m}=\left(  \kappa\frac{\ddot
{a}_{m}}{\dot{a}_{m}}\right)  \left(  \frac{\dot{a}_{m}}{a_{m}}\right)
=\left(  \kappa\frac{-\left(  \frac{2\pi}{\tau}\right)  ^{2}a_{\max}%
\sin\left(  2\pi\frac{t}{\tau}+\psi\right)  }{\frac{2\pi}{\tau}a_{\max}%
\cos\left(  2\pi\frac{t}{\tau}+\psi\right)  }\right)  \left(  \frac{\frac
{2\pi}{\tau}a_{\max}\cos\left(  2\pi\frac{t}{\tau}+\psi\right)  }{a_{\max}%
\sin\left(  2\pi\frac{t}{\tau}+\psi\right)  }\right)  =-\left(  \frac{2\pi
}{\tau}\right)  ^{2}\kappa$. For $\kappa=-3$, we verify that$\ \frac{1}%
{3}\left\vert \Gamma\right\vert H_{m}=\omega^{2}$, justifying transforming of
Eq. 4.1 into Eq. 4.2. We define a cosmic period parameter $\tau$ in the model as,%

\begin{equation}
\tau=2\pi\sqrt{\frac{3}{\left\vert \Gamma\right\vert H_{m}}}. \tag{4.4}%
\end{equation}

\section{Dynamics of a self-regulating universe}

We have proposed an approach embodied in three propositions, in which the two
ingredients can be brought together into a framework that justifies the
process of gravitationally-induced matter creation/annihilation in dynamics of
a universe in this model. In this Section we put together the results into a
combined framework based on the Dynamic Protection Proposal (DEPP).

\subsection{ Combining the dynamics}

Here we seek to describe the expansion rate $a(t)$ of this universe implied by
combining the results in Eqs. 2.9 and 4.3. We note that while the two
contributions to the expansion are separately sourced, as by implication of
the Cosmological Principle, their expansion effects are collinear. The
resulting time evolution of the scale-factor $a(t)$\ is therefore an algebraic
sum of the contributions, $a(t)=a_{s}(t)+a_{m}(t)$, consistent with the
expectations of Eq. 2.11 provided we set $\delta a_{s}=a_{m}$. This then gives
the general evolution of the scale-factor as
\begin{equation}
a(t)=\upsilon^{\ast}t+a_{\max}\sin\omega t+\psi\tag{5.1}%
\end{equation}
Further, the general expression for the Hubble parameter, $H\left(  t\right)
=\frac{\dot{a}}{a}$ can be written down directly from Eq. 5.1 as%

\begin{equation}
H(t)=\frac{\upsilon^{\ast}+a_{\max}\omega\cos\left(  \omega t+\psi\right)
}{\upsilon^{\ast}t+a_{\max}\sin\left(  \omega t+\psi\right)  }. \tag{5.2}%
\end{equation}
The results obtained in Eq. 5.1 and Eq. 5.2 imply the merging of the two
ingredients of the approach.

\subsection{Boundary and initial conditions}

In order that the two ingredients carried by Eq. 2.9 and Eq. 4.3 are linkable
to represent the dynamics of the same physical system, the universe, it is
desirable that the differential equation of Eq. 4.2 be subject to appropriate
boundary conditions. This includes fixing both the oscillation amplitude
$a_{\max}$ and the phase angle $\psi$. Below we briefly discuss the rationale
for the appropriate choice in fixing each.

\subsubsection{The "no-stalling" condition}

The contribution $a_{m}(t)$ from matter-energy sector in Eq. 4.3 to the scale
factor $a\left(  t\right)  $ in Eq. 5.1 constitutes a perturbation on a
time-linearly expanding background $a_{s}$. Recall, by Proposition 2 the
universe, in the model, seeks to maintain cosmic dynamics stable to
perturbations. In the process it creates matter-energy, whose effect is
manifested by the $a_{m}(t)$ perturbative contribution from Eq. 4.3. In order
that these do not grow into run-away perturbations, it is important that
$\dot{a}(t)>0$ for all $t$ values. Thus if we denote by $\dot{a}_{\min}$, the
minimum value of the net expansion rate $\dot{a}\left(  t\right)
=\upsilon^{\ast}+a_{\max}\omega\cos\left(  \omega t+\psi\right)  $, then
noting that such a minimum occurs at $\omega t+\psi=\pi(2n+1),n=0,1,2...$, we
can write $\dot{a}_{\min}=\upsilon^{\ast}-a_{\max}\omega=\alpha$, where the
parameter $\alpha$ can be arbitrarily small but satisfies$\ \alpha>0$. We
refer to this constraint as the no-stalling condition. In terms of
$\upsilon^{\ast}$ one can write this as $\dot{a}_{\min}=\upsilon^{\ast
}-a_{\max}\omega=\upsilon^{\ast}\beta$, where we shall later show
$\beta\gtrsim$ $0$. From this we therefore have as a boundary condition that
Eq. 4.2 is subject to%

\begin{equation}
a_{\max}=\upsilon^{\ast}\tilde{\tau}\left[  1-\beta\right]  , \tag{5.3}%
\end{equation}
where, $\tilde{\tau}=\frac{\tau}{2\pi}=\frac{1}{\omega}\ $is the\textit{
reduced period }(a notation we shall often use for brevity). Also for brevity
we shall often write $\left[  1-\beta\right]  =\eta$, where (see Section
5.2.2) $\eta\lesssim1$. Use of Eq. 5.3 in Eqs. 5.1 and 5.2 gives the
evolutions of the scale factor and Hubble parameter, respectively, as%

\begin{align}
a(t)  &  =\upsilon^{\ast}\left[  t+\tilde{\tau}\eta\sin\left(  \frac{t}%
{\tilde{\tau}}+\psi\right)  \right]  ,\tag{5.4a}\\
H(t)  &  =\frac{1+\eta\cos\left(  \frac{t}{\tilde{\tau}}+\psi\right)
}{t+\tilde{\tau}\eta\sin\left(  \frac{t}{\tilde{\tau}}+\psi\right)  }.
\tag{5.4b}%
\end{align}

Recall previously in Eq. 2.9 we set $\upsilon^{\ast}=\frac{1}{t_{0}}$, for the
idealized linear case $a_{s}=\upsilon^{\ast}t$. We shall now re-calibrate
$\upsilon^{\ast}$ (for convenience) and generally define it as $\upsilon
^{\ast}=\frac{1}{t_{0}+\tilde{\tau}\eta\sin\left(  \frac{t_{0}}{\tilde{\tau}%
}+\psi\right)  }$, where $t_{0}$ is still the current age of the Universe, so
that $a_{0}=1$. Note, here, we generally still have.$\upsilon^{\ast}=\frac
{1}{t_{0}}$ at $a(t)=a_{s}(t)$, whenever $\omega t+\psi=n\pi,n=0,1,2,..$

Finally, as can be verified by inspection, Eq. 5.1 satisfies a second order
differential equation of the form
\begin{equation}
\tilde{\tau}^{2}\ddot{a}+\left(  a-\upsilon^{\ast}t\right)  =0. \tag{5.5}%
\end{equation}
Therefore, Eq. 5.5 will be identified as the equation of motion that describes
cosmic dynamics in this model, with Eqs. 5.4 as its solutions.

\subsubsection{\textbf{The phase angle }$\psi$}

From Eq. 5.1 we have at $t=0$ that $a_{t=0}=a_{\max}\sin\psi$ and that
$\dot{a}_{t=0}=\upsilon^{\ast}+a_{\max}\omega\cos\psi$, which on use of Eq.
5.3 give%

\begin{align}
a_{t=0}  &  =\upsilon^{\ast}\tilde{\tau}\eta\sin\psi\tag{5.6a}\\[0.03in]
\dot{a}_{t=0}  &  =\upsilon^{\ast}\left(  1+\eta\cos\psi\right) \tag{5.6b}\\
H_{t=0}  &  =\left(  \frac{\dot{a}}{a}\right)  _{t=0}=\frac{1+\eta\cos\psi
}{\tilde{\tau}\eta\sin\psi}. \tag{5.6c}%
\end{align}
Eqs. 5.6 constitute a statement on initial conditions of cosmic expansion
parameters in this approach. These results imply that, provided $\psi\neq0$,
the state the model universe emerges from at $t=0$ has finite and
non-vanishing geometrical parameters. In section 5.2.4 we discuss more on
initial conditions.

We now share some thoughts on how one may constrain $\psi$. Recall the above 3
attributes imply he universe is perpetually in search for its dynamic
equilibrium state of a \textit{specific} time-linear expansion $a_{eq}%
(t)=a_{s}(t)=\upsilon^{\ast}t$. It follows, from Eq. 5.6b, that, the initial,
state for the Universe, $t=0$, must be a maximally out of equilibrium state,
with regard to its expansion rate, so that $\dot{a}_{t=0}=\upsilon^{\ast
}\left(  1+\eta\cos\psi\right)  =\left(  \frac{da}{dt}\right)  _{\max}$. The
maximal deviation of the expansion rate from equilibrium at $t=0$, is $\left(
\delta\dot{a}\right)  _{\max}=\left[  \dot{a}_{t=0}-\dot{a}_{eq}\right]
=\left[  \upsilon^{\ast}\left(  1+\eta\cos\psi\right)  -\upsilon^{\ast
}\right]  $. Giving%

\begin{equation}
\left(  \delta\dot{a}\right)  _{\max}=\upsilon^{\ast}\eta\cos\psi\tag{5.7}%
\end{equation}
In order for $\left(  \delta\dot{a}\right)  _{\max}$ to be as large as
possible one must have $\psi\left(  <\frac{\pi}{2}\right)  \ $to be as small
as possible.

Now in building $\psi$ the only parameters considerable in the model are: (i)
the time scale $\tau$ and (ii) some fundamental length scale $l_{f}$ (such as
the Planck length $l_{f}\sim l_{p}=\left(  \frac{\hslash G}{c^{3}}\right)
^{\frac{1}{2}}$. One can build from these two parameters a vanishingly small
angle by setting $\psi\sim\frac{l_{f}}{c\tau}$, where $c$ is the velocity of
light. In terms of the scale-factor $a\left(  t\right)  $, we have $\psi
\sim\frac{l_{f}}{c\tau}=\frac{a_{l_{f}}}{\upsilon^{\ast}\tau}$, where
$a_{l_{f}}$ is the scale-factor associate with $l_{f}$ and $\upsilon^{\ast
}\tau$ is the scale factor at $t=\tau$, so that $\psi\sim\frac{l_{f}}{c\tau
}=\frac{a_{l_{f}}}{\upsilon^{\ast}\tau}$. Thus, considering the $a_{t=0}$
given in Eq. 5.6a and substituting for $\psi$ gives, $a_{t=0}=\upsilon^{\ast
}\tilde{\tau}\eta\sin\left(  \frac{a_{l_{f}}}{\upsilon^{\ast}\tau}\right)  $,
which, in the small angle limit, $\lim\limits_{\psi\rightarrow0}\sin\psi=\psi$
(and $\eta\simeq1$) yields (on use of $\eta\simeq1$) $a_{t=0}=a_{_{l_{f}}}$,
Therefore the initial size of the universe in the model can be set to
$R_{t=0}=l_{f}$ provided one sets $\psi\sim\frac{l_{f}}{c\tau}$.

\subsubsection{Constraints on $\eta$}

The solutions in Eqs. 5.4 imply the model has 3 free parameters $\beta$,
$\psi$ and $\tau$. In the preceding discussion we showed that $\psi$ can be
vanishingly small and we briefly discussed on how it can be constrained. From
the discussion leading to Eq. 5.3 recall the no-stalling condition implies
$\dot{a}_{\min}=\upsilon^{\ast}\beta>0$. If we were to set $\beta
=0\Rightarrow\eta=1$, then $\dot{a}_{\min}=0$ and the Universe stalls. Since
$\beta$ can be made arbitrarily small, we would like to set appropriate
conditions $\beta=\beta\left(  \psi\right)  $ such that $\beta$ is as small as
possible but $\beta>0$. This has the advantages that (i) the free parameters
in the model are reduced to only 2, and (ii) $\psi$ now determines, not just
the initial conditions in Eqs. 5.6, but also regulates the dynamics of the
Universe from stalling. These characteristics are obtained if\ we set
$\beta=\sin\psi$, which also recall $\sim\psi$. A detailed discussion of this
feature and the implications of the no-stalling condition on cosmic dynamics
will appear in an up-coming piece of work.

\subsubsection{On horizon problem and flatness problems}

The model can be tested for both the horizon and flatness problems. From Eq.
5.4a it is clear that $\int_{0}^{t}\frac{dt}{a}=\int_{0}^{t}\frac{dt}%
{\upsilon^{\ast}\left[  t+\tilde{\tau}\eta\sin\left(  \frac{t}{\tilde{\tau}%
}+\psi\right)  \right]  }$ diverges at $t=0$. Thus all the space then is
within the horizon and the model has no horizon problem. Further from Eq. 5
one notes that at $t=0$, $\dot{a}=\upsilon^{\ast}\left[  1+\eta\cos\left(
\frac{t}{\tilde{\tau}}+\psi\right)  \right]  $ has no initial singularity.
Consequently, at $t=0$ the model has no flatness problem either.

\section{Consistency with cosmological observations}

The effectiveness of the framework just set up to describe a working cosmic
dynamics can now be tested. Here, we seek to establish the extent to which the
framework reproduces the known profile of evolutionary features of the
Universe in general, and whether in particular, it recovers the observed
values of the standard cosmological parameters. In discussing these
macroscopic features, we shall, with no loss of generality, set $\psi=0$. Then
from Eqs. 5.4 the working equations for cosmic dynamics here become,
\begin{align}
a(t)  &  =\upsilon^{\ast}\left[  t+\tilde{\tau}\sin\frac{t}{\tilde{\tau}%
}\right]  ,\tag{6.1a}\\
H(t)  &  =\frac{1+\cos\frac{t}{\tilde{\tau}}}{t+\tilde{\tau}\sin\frac
{t}{\tilde{\tau}}}. \tag{6.1b}%
\end{align}

\subsection{Age $t_{0}$, period $\tau$ and Hubble constant $H_{0}$}

Here we calculate cosmological parameters from the model and compare the
results with known values. Observations [10] show that the product of current
physical age of the Universe $t_{0}$ and the Hubble constant $H_{0}$, referred
to as the \textit{dimensionless age} is currently\footnote{On a side note,
this result that currently $H_{0}t_{0}\approx1$, actually leads to a puzzle
which has recently been referred to as the \textit{Synchronicity Problem}
[10], and which we will later seek to address, separately.} $H_{0}t_{0}%
\approx1$. This estimate when applied on Eq. 6.1b implies, in our case, we can
write $H_{0}t_{0}=\left(  1+\cos\frac{t_{0}}{\tilde{\tau}}\right)  \left(
t_{0}+\tilde{\tau}\sin\frac{t_{0}}{\tilde{\tau}}\right)  ^{-1}t_{0}\approx1$.
More specifically, observations\ from type Ia supernovae (SN), combined with
those from baryon acoustic oscillations (BAO) and from cosmic microwave
radiation (CMB) data [37] constrain the dimensionless age\footnote{It should
be mentioned that, while this is the tightest constraint available currently,
it is also potentially $\Lambda CDM$ model-dependent.} to $H_{0}t_{0}=0.96$.

Additionally, the look-back time when the currently observed cosmic
acceleration started, and which we call here the cosmic acceleration
commencements look-back time $t_{ac}$, is estimated [3] to be around 4 billion
years. From the established chronology of cosmic evolutionary phases [4] we
shall take the current acceleration to be the first since the primordial
acceleration [5] so that $\tau_{\frac{1}{2}}<t_{0}<\tau$, where $\tau
_{\frac{1}{2}}=\frac{1}{2}\tau$. We can then rewrite the cosmic period in the
model as $\tau_{\frac{1}{2}}=t_{0}-t_{ac}$, where $t_{0}$ is the age so that
the above dimensionless age $H_{0}t_{0}$ expression becomes
\begin{equation}
\frac{1+\cos\left(  \frac{\pi t_{0}}{t_{0}-t_{ac}}\right)  }{1+\left(
\frac{t_{0}-t_{ac}}{\pi t_{0}}\right)  \sin\left(  \frac{\pi t_{0}}%
{t_{0}-t_{ac}}\right)  }\approx0.96.\tag{6.2}%
\end{equation}
Therefore provided we can constrain $t_{ac}$ to the around the accepted
estimate of $t_{ac}\sim4Gyr$ [3] we can use Eq. 6.2 to estimate, in this
model, the age $t_{0}$, the period $\tau=2\left(  t_{0}-t_{ac}\right)  $, and
along with use Eq. 6.2b we also obtain $H_{0}$. Table 1 below gives the
calculated values of $t_{0}$, $\tau_{\frac{1}{2}}$, $H_{0}$ for a range of
values of $4.02\leq t_{ac}\leq4.12$. It also gives a range of values (to be
discussed in next subsection 6.2) for the synchronicity time $t_{sy}$ which
gives $Ht_{sy}=1$
\begin{align*}
&
\begin{tabular}
[c]{|c|c|c|c|c|}\hline
$Input$ & \multicolumn{4}{|c|}{$Output$}\\\hline
$t_{ac}/Gyr$ & $t_{0}/Gyr$ & $H_{0}/kms^{-1}/Mpc$ & $\tau/Gyr$ & $t_{sy}%
/Gyr$\\\hline
$4.02$ & $13.59$ & $69.08$ & $19.14$ & $13.69$\\\hline
$4.04$ & $13.66$ & $68.697$ & $19.24$ & $13.76$\\\hline
$4.06$ & $13.73$ & $68.40$ & $19.33$ & $13.83$\\\hline
$4.07$ & $13.76$ & $68.22$ & $19.38$ & $13.86$\\\hline
$4.08$ & $13.79$ & $68.11$ & $19.42$ & $13.89$\\\hline
$4.10$ & $13.86$ & $67.739$ & $19.52$ & $19.96$\\\hline
$4.12$ & $13.93$ & $67.37$ & $19.62$ & $14.03$\\\hline
\end{tabular}
\\
Table1 &  :Estimates\ of\ some\ time\ based\ cosmological\text{\ }%
parameters\ for\ range\ 4.02\leq t_{ac}\leq4.12
\end{align*}
One notes from Table 1, that for the range taken the corresponding results of
cosmic age, $t_{0}$ is within cosmic age estimates (12-15Gyr) based on oldest
stars from globular clusters [38]. From this range $4.02\leq t_{ac}%
\leq4.12Gyr$ we find that $t_{ac}=4.07$ $Gyr$ gives the combined results of
$t_{0}$ and $H_{0}$ most consistent with latest observations. We comment that
this result is not a result of fine-tuning. The model depends on observational
constraints on $t_{ac}$ used in Eq. 6.2 and hence use the range $4.02\leq
t_{ac}\leq4.12Gyr$, which clearly is quite tight. We also used the range
extent $4.02-4.12$\ $Gyr$ to generate (only range-based theoretical) error
bars on the results. Thus, for $t_{ac}=4.07\pm0.05$ $Gyr\ $we find (see Table
1) as an estimate of the age of the Universe $t_{0}=13.76_{-0.17}^{+0.17}Gyr$.
Besides lying within the age-range given above by globular clusters [38], this
result further agrees very well with latest tighter observational constraints
[6].\ We also find (see Table 1) the cosmic period parameter as $\tau
=19.38_{-0.24}^{+0.24}Gyr$. We observe that this latter parameter $\tau$ is,
to our knowledge a new parameter, specific to the model. Finally, use of Eq.
6.1b gives our working estimate (see Table 1) for Hubble constant as
$H_{0}=68.22_{-0.85}^{+0.87}\pm kms^{-1}/Mpc$. This result agrees well with
recent observations, particularly showing little or no statistical difference
from those by Freedman (2021) [39] of $H_{0}=69.8\ \pm.06(stat)\pm
1.6(sys)\ kms^{-1}/Mpc\ $and those by N. Khetan (2021) [40], et al of
$70.50\pm2.37(stat.)\pm3.38\ (sys.)\ kms-1Mpc-1$\textbf{.}

Figure 1 is a diagrammatic sketch of the time evolution of the cosmic
scale-factor $a(t)$ of Eq. 5.9a based on $\tau_{\frac{1}{2}}\simeq9.69$
$Gyrs$. It shows a harmonic variation of $a(t)$ (red curve) about the dynamic
equilibrium (or coasting) expansion path $a_{s}=\upsilon^{\ast}t$ (blue line)
derived in Eq. 2.9. The $a(t)$ curve here clearly displays the features
embodied in the DEPP Proposal (Section 4.1). It shows that the ongoing
acceleration which began at look back time $t_{ac}=4.07$ $Gyr$ (i.e.
$t=\tau_{\frac{1}{2}}\simeq9.69Gyrs)$ is currently ongoing (consistent with
observations) and, according to the model, ends at \textbf{cosmic age }%
$t=\tau\simeq19.38Gyrs$\textbf{ \ }In Appendix 1 we show a diagramatic sketch
comparing cosmic evolution in this non-exponetial acceleration model and that
of inflationary models, such as $\Lambda CDM$.%

\begin{figure}
\centering
\includegraphics[scale=1.0]{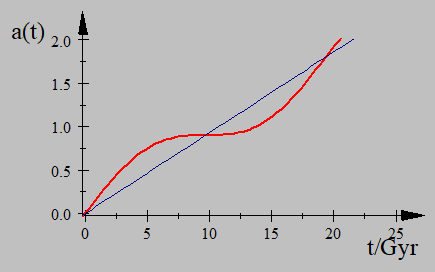}
\caption{\textit{Shows the general features of the time evolution of the
scale factor }$a(t)$\textit{, as depicted by the model. The net expansion
(red) which is a combination of two independent influences (basic-space and
free matter-energy) is shown as an oscillatory curve about the linear
(equilibrium) expansion rate of basic-space. Here we have used }$\tau
_{\frac{1}{2}}=9.69Gyr$\textit{\ as the working cosmic half period. Note:
}$a_{0}\ =1$ at $t=13.76Gyr$\textit{.}}%
\label{Figure 1.png}%
\end{figure}

Figure 2 shows the evolution of the Hubble parameter (red curve) of Eq. 5.9b
based on $\tau_{\frac{1}{2}}=9.69$ $Gyr$. The curve oscillates about the
would-be equilibrium (or coasting) case (blue hyperbola). Note that here the
minimum value of the expansion rate first occurs at same time $\tau_{\frac
{1}{2}}=9.69$ $Gyr$ when also (in Figure 1) the general $a(t)\ $curve crosses
the equilibrium path-line$\ a_{s}=\upsilon^{\ast}t$. Thus, it is at
$\tau_{\frac{1}{2}}=9.69$ $Gyr$ the Universe first$\ $virtually comes closest
to a halt and also when the current cosmic acceleration commences.%

\begin{figure}
\centering
\includegraphics[scale=1.0]{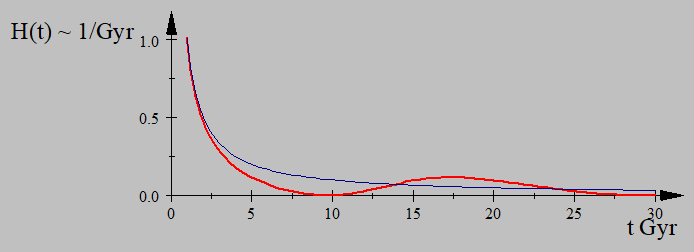}
\caption{\textit{The time evolution of the Hubble parameter }$H(t)\ $(red
curve)\textit{.} \textit{The curve is a periodic oscillation, period},\textit{
about the would-be equilibrium evolutionary path of constant expansion. The
intersections between the }$H(t)$\textit{ curve and the }$\frac{1}{t}$\textit{
hyperbola identify the "synchronous points" where }$H(t)t=1\ $(see discussion
below)\textit{.}}%
\label{Figure 2.png}%
\end{figure}

\subsubsection{The dimensionless age \textit{synchronicity }problem: why is
$H_{0}t_{0}\simeq1$ now$?$}

We now discuss the dimensionless age puzzle.

\textbf{Problem definition: }Cosmological observations, including CMB power
spectrum, baryon acoustic oscillations (BAO) and type Ia supernovae (SN Ia)
apparent magnitude and redshift measurements [10] [37] do indicate that the
current dimensionless age of the Universe appears to be close to unity,
$H_{0}t_{0}\simeq1$. To date, this observation has no satisfactory
explanation. When this product is exact unity, it implies the average rate of
expansion of the Universe over its entire age,$\frac{\Delta a}{\Delta t}%
=\frac{a}{t}$, is exactly the same as the Universe's current local rate of
expansion $\dot{a}\left(  t\right)  $. However, given a history of varied
cosmic expansion rate, based on different past cosmic phases, such a result
appears puzzling. The \textit{Synchronicity Problem} [10] then arises out of
lack of a satisfactory explanation of the suggestion, implied by this result,
that we currently live in a special era when the two ratios are virtually
equal. A related question is why the observed current cosmic state gives a
result [37] so close to, but not exactly, unity either.

\textbf{Analysis and resolution: }In our approach $H\left(  t\right)  t=1$ has
a natural explanation which also leads to a resolution of the
\textit{Synchronicity Problem}. Recall from Eq. 6.1a the scale-factor $a(t)$
evolves as a periodic function about a coasting path $\upsilon^{\ast}t$. Let
us define the "\textit{synchronous time}" $t=t_{sy}$\ as the time, measured
from $t=0$, when in general the relation $\left(  \frac{\Delta a}{\Delta
t}=\frac{a}{t}\right)  =\dot{a}\left(  t\right)  \Longrightarrow H\left(
t\right)  t=1$ is satisfied. Then at $\ t=t_{sy}$ we have in our case, on use
of Eq. 6.1a, that $\frac{\upsilon^{\ast}\left[  1+\cos\left(  \frac{t_{sy}%
}{\tilde{\tau}}\right)  \right]  t_{sy}}{\upsilon^{\ast}\left[  t_{sy}%
+\tilde{\tau}\sin\left(  \frac{t_{sy}}{\tilde{\tau}}\right)  \right]  }=1$.
This sets the "synchronicity" condition in this approach, as%

\begin{equation}
\frac{\tilde{\tau}_{\frac{1}{2}}}{t_{sy}}\tan\left(  \frac{t_{sy}}{\tilde
{\tau}_{\frac{1}{2}}}\right)  =1, \tag{6.3}%
\end{equation}
where, as before, $\tilde{\tau}_{\frac{1}{2}}=\frac{\tau_{\frac{1}{2}}}{\pi
}=\frac{\tau}{2\pi}=\tilde{\tau}$ is the reduced half cosmic period. Note that
Eq. 6.3 and Eq. 6.2 refer to, closely related but, different scenarios. Eq.
6.3 admits periodic solutions of the form
\begin{equation}
t_{sy-n}=t_{sy-1}+(n-1)\tau_{\frac{1}{2}}, \tag{6.4}%
\end{equation}
with $n=1,2....$. Using the previously calculated value of $\tau=2\tau
_{\frac{1}{2}}=19.38$ $Gyr$ (see also Table 3) we find $t_{sy-1}%
=13.86_{-0.35}^{+0.17}$ $Gyr$. From Eq. 6.4 the next \textit{synchronous
}point \textit{at} $t=t_{sy-2}$ appears as $t_{sy-2}=23.55\ Gyr$.

Figure 3 demonstrates the synchronicity situation. The \textit{synchronous
}point \textit{at} $t=t_{sy-1}$ is shown where the (\textit{magenta}) line
from origin is tangent to the $a\left(  t\right)  $ (red) curve. As can be
seen from Figure 3 this point coincidentally happens to fall within the (model
independent ) observational limits [35] of the current age of the Universe
$12-15$ $Gyr$.%
\begin{figure}
\centering
\includegraphics[scale=1.2]{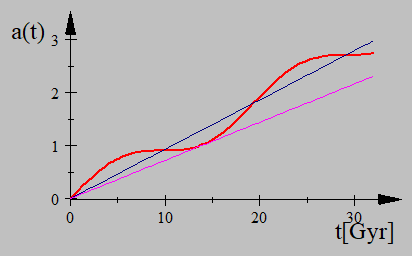}
\caption{\textit{Time evolution of }$a(t)$\textit{ (similar to Figure 1). The
\textbf{synchronous} point appears where the purple line is tangent to the
curve where }$H\left(  t\right)  t=1$ (\textit{See also Figures 2 3 and 4).}}%
\label{Figure3.png}%
\end{figure}

Figures 2 and 4 identify the \textit{synchronous }points $t_{sy-n}$ much more
clearly. In each case, the points appear, respectively, at the intersections
of the $H(t)$ curve and the $\frac{1}{t}$ hyperbola, and at the intersection
of\textit{\ the }$H(t)t$ curve and $f(t)=1$.%

\begin{figure}
\centering
\includegraphics[scale=1.2]{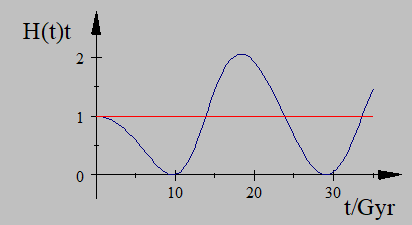}
\caption{\textit{Plot of dimensionless age H(t)t as a function of time t in
the model. The Synchronous points appear where this function intersects
}$f(t)=1\ $\textit{(blue). They correspond to the points of intersection in
Figure 2. Note here that }$t=0=t_{sy-0}$ \textit{is the first of synchronous
points. The Universe is currently approaching the second one.
\ \ \ \ \ \ \ \ \ \ \ }}%
\label{Figure 4.png}%
\end{figure}

Recall we previously estimated the age of the Universe as $t_{0}=13.76\ Gyr$.
Inspection of Figures 2 and 3 shows that currently $t_{0}<t_{sy-1}$ and the
expansion rate $\dot{a}\left(  t\right)  \mid_{t_{0}}$ is very close to, but
not quite, that at synchronous time, $\ \dot{a}\left(  t\right)  \mid_{t_{0}%
}\lesssim\dot{a}\left(  t\right)  \mid_{t_{sy}}$. According to our current
estimate cosmic dynamics will therefore soon become synchronous, at a time
from now, of $\left(  t_{sy}-t_{0}\right)  \simeq\left(  13.86-13.76\right)
=0.10$ $Gyr$. This is only $\sim100$ million years from now. As Figures 2, 3-
and 4, this event is a \textit{once-in-a-half-cosmic-cycle} event, $\Delta
t=\tau_{\frac{1}{2}}$. Therefore the implication by this analysis is that,
according to this model, we currently live in a special era, very close to the
first of these special events when the age of the Universe and the Hubble
constant become synchronized.

\subsection{Density parameters and their evolution}

We close this section with a brief discussion regarding the matter-energy
fields in the model and their evolution. The net gravitational mass density of
all the matter-fields driving the dynamics, in the model, can be inferred from
the relation $\frac{\ddot{a}}{a}+\frac{4\pi G}{3}\left(  \rho+3p\right)  =0$
of Eq. 2.4. Using the solution $a\left(  t\right)  =a(t)=\upsilon^{\ast
}\left[  t+\tilde{\tau}\sin\frac{t}{\tilde{\tau}}\right]  $ from Eq. 6.1a we
find on denoting, here, $\rho+3p=\Delta$, that the net gravitational mass
density evolves as
\begin{equation}
\Delta=\frac{3}{4\pi G}\left[  \frac{\frac{t}{\tilde{\tau}}\sin\frac{t}%
{\tilde{\tau}}+\sin^{2}\frac{t}{\tilde{\tau}}}{a^{2}}\right]  . \tag{6.5}%
\end{equation}
We note that this is made up of two fields, referred to here as $\Delta_{++}$
and $\Delta_{+-}$with different evolutionary characteristics, such that
$\Delta=\Delta_{++}+\Delta_{+-}$, with%
\begin{equation}
\Delta_{++}=\frac{3}{4\pi G}\left[  \frac{\sin^{2}\frac{t}{\tilde{\tau}}%
}{a^{2}}\right]  , \tag{6.6}%
\end{equation}
and
\begin{equation}
\Delta_{+-}=\frac{3}{4\pi G}\left[  \frac{\frac{t}{\tilde{\tau}}\sin\frac
{t}{\tilde{\tau}}}{a^{2}}\right]  . \tag{6.7}%
\end{equation}

\subsubsection{Some effective features of the fields $\Delta_{+-}$ and
$\Delta_{++}:\ $Cosmic acceleration and flat rotation curves}

The gravitational mass density $\Delta$ of Eq. 6.5 represents the net
gravitating fields in the model. As Eqs. 6.6 and 6.7 indicate $\Delta$ is made
up of two kinds of fields $\Delta_{++}$ and $\Delta_{+-}$ which are
functionally different. Below we highlight on some of these characteristics.

\textbf{Features of }$\Delta_{+-}$\textbf{ as dark energy: }The field
$\Delta_{+-}$ is an odd function and changes its gravitational charge
signature periodically, being currently negative. We will take $\Delta_{+-}$to
contribute the net negative pressure component of dark energy which provides
the acceleration of the universe in the model. To justify this assumption we
evaluate $\Delta_{+-}$ in relation to $\Delta_{++}$, in section 6.2.2 below,
and compare the results with observations.

\textbf{Features of }$\Delta_{++}$ as\textbf{ matter/dark matter: }The field
$\Delta_{++}$ is an even function with positive-definite gravitational charge
signature, reminiscent of regular or dark matter. Because in this model,
fields are only distinguished by their net gravitational charge, $\Delta_{++}$
represents all matter including dark matter, baryonic matter, radiation and
neutrinos (and excludes negative pressure\footnote{In this model it is not
necessary to invoke negative energy densities. Just as in the $\Lambda CDM$
model, it is sufficient to have negative pressure.} included in $\Delta_{+-}%
$). For a realistic model consistent with observations, one expects
$\Delta_{++}$ to be dominated by a dark energy density $\rho_{d}$. To test
this we make the reasonable assumption, based on its positive gravitational
signature, that during structure formation the field $\Delta_{++}$ coalesces
into galaxies enveloped, and dominated, by dark matter haloes. Considering
$M\left(  R\right)  $ as the mass of such a halo enclosed by a distance $R$
from the origin, where $R>R_{0\text{ }}$($R_{0\text{ }}$ being the extent of
the baryonic matter in the galaxy, of mass $M_{0}$), then along $R>R_{0\text{
}}$, the orbital velocities $v$ of the galaxy satellites can be obtained from
their centripetal and gravitational accelerations relation, $\frac{v^{2}}%
{R}=\frac{GM}{R^{2}}$. In our case $\frac{GM}{R^{2}}\sim\frac{4}{3}\pi
GR\Delta_{++}$. Given (see Eq. 6.6) that $\Delta_{++}\propto\frac{\sin
^{2}\frac{t}{\tilde{\tau}}}{a^{2}}\propto\frac{\sin^{2}\frac{t}{\tilde{\tau}}%
}{R^{2}}$, and noting that $\sin^{2}\frac{t}{\tilde{\tau}}$ is a slowly
varying function (with a long period $\tau\approx19.38Gyr$), we see that for
distances of few tens $kpcs$ usually associated with such halos [41] $v$ is
constant. It follows then that the model produces fields identifiable as dark
matter and whose halos have the physical structure to generate flat rotation
curves in galaxies. Such flat rotation curves are consistent with observations
[42] [43] [44]. We therefore consider this a noteworthy result of the model,
whose detailed analysis and discussion will follow in an upcoming work.

\subsubsection{Comparison with observations}

It is reasonable to ask how the above gravitational mass density results do
relate to current observations of an accelerating Universe. In principle, one
should relate the strength (density) of the negative pressure causing the
acceleration of space to the available inertia from the matter-energy density.
In our case, the key quantities to compare with observations are the net
accelerating field density $\Delta_{+-}$ and the source of inertia to be
accelerated $\Delta_{++}$, and the comparison can sufficiently be expressed in
the form $\frac{\Delta+-}{\Delta++}$.

In the $\Lambda CDM$ model, cosmic acceleration is assumed to be driven by a
dark energy with an energy density $\rho_{de}$ and pressure $p_{de}$ which can
be represented as an energy momentum tensor of a perfect fluid $T^{\mu\nu
}=diag(\rho_{de},p_{de},\rho_{pde},p_{de})$. Its contribution to cosmic
dynamics can be inferred from Eq. 2.4 as $\frac{\ddot{a}}{a}+\frac{4\pi G}%
{3}\left(  \rho_{de}+3p_{de}\right)  =0$. In particular, in $\Lambda CDM$ the
dark energy density $\rho_{de}$ is assumed constant in time, satisfying an
equation of state $p_{de}=-\rho_{de}$. It follows that, for a field with an
energy density $\rho_{de}$, the driver of cosmic acceleration in $\Lambda CDM$
is a net negative pressure equivalent to $\rho_{de}-3\rho_{de}=-2\rho_{de}$.
Therefore, from observations the ratio of the source of the acceleration to
the source of inertia $\rho_{m}$ would be $\frac{\left\vert -2\rho
_{de}\right\vert }{\rho_{m}}$. This result $\frac{\left\vert -2\rho
_{de}\right\vert }{\rho_{m}}$, just like $\frac{\left\vert \Delta
_{++}\right\vert }{\Delta_{+-}}$, is model-independent. In our present case,
using Eq. 6.6 and Eq. 6.7 and our previously determined values of
$t_{0}=13.76Gyrs$ and $\tau\simeq19.38Gyrs$ we find $\frac{\left\vert
\Delta+-\right\vert }{\Delta++}=4.\,605\,2$. Below, we show that this result
is consistent with recent observations [6] (usually represented differently).

Traditionally, the results giving the ratio of the source of the acceleration
to the inertia to accelerate is given, instead, in terms of the dark energy
density $\rho_{de}$ and the matter-energy density $\rho_{m}$ (or more
precisely in terms of their weighted parameters $\Omega_{de}$ and $\Omega_{m}%
$). This follows the observation that in $\Lambda CDM$, the driving pressure
and the dark energy density are connected by the equation of state
$p_{de}=-\rho_{de}$, making it more intuitively easier to compare the
densities, instead. While valid for $\Lambda CDM$, such presentation is,
clearly, model-dependent.

For easier, more transparent, verification we will now present the above
result ($\frac{\left\vert \Delta+-\right\vert }{\Delta++}=4.\,605\,2$) using
the (now common $\Lambda CDM$ ) approach of the density $\rho_{de}$ instead of
the driving pressure $-2\rho_{de}$. To do this, denote by $\left\vert
\Delta_{\frac{1}{2}\left(  +-\right)  }\right\vert =\frac{1}{2}\left\vert
\Delta_{+-}\right\vert $ the analogue of $\rho_{de}=\frac{1}{2}\left\vert
2p_{de}\right\vert $ in $\Lambda CDM$. Then in our model, the analogue of the
net density in $\Lambda CDM$ would be $\left\vert \Delta_{\frac{1}{2}\left(
+-\right)  }\right\vert +\Delta_{++}$ It follows that the percentage
contributions to this by the dark energy, and by matter, in our model at any
time $t$ can be respectively computed from $\frac{\left\vert \Delta_{\frac
{1}{2}\left(  +-\right)  }\right\vert }{\left\vert \Delta_{\frac{1}{2}\left(
+-\right)  }\right\vert +\Delta_{++}}$ and $\frac{\Delta_{++}}{\left\vert
\Delta_{\frac{1}{2}\left(  +-\right)  }\right\vert +\Delta_{++}}$. In
particular one finds the current values to be $\Omega_{de}=\left[
\frac{\left\vert \Delta_{\frac{1}{2}\left(  +-\right)  }\right\vert
}{\left\vert \Delta_{\frac{1}{2}\left(  +-\right)  }\right\vert +\Delta_{++}%
}\right]  _{t=t_{0}}=0.697\,21$ and $\Omega_{m}=\left[  \frac{\Delta_{++}%
}{\left\vert \Delta_{\frac{1}{2}\left(  +-\right)  }\right\vert +\Delta_{++}%
}\right]  _{t=t_{0}}=0.302\,79$, where we used our results of $t_{0}=13.76Gyr$
and $\tau=19.38Gyr$.

Therefore summarizing, we have that taking the field $\Delta_{+-}$ in Eq. 6.6
to constitute the net driver of the acceleration (with $\left\vert
\Delta_{\frac{1}{2}\left(  +-\right)  }\right\vert $ as its effective density
corresponding to $\rho_{de}$ in $\Lambda CDM$) and taking the field
$\Delta_{++}$ in Eq. 6.7 to constitute the mass-energy density (analogous to
$\rho_{m}$) we have shown that, in this model, the analogous dark energy
density contributes $\Omega_{de}=69.7\%$ while the entire matter sector
contributes $\Omega_{m}=30.3\%$. These results are consistent with the recent
observations [6].

\subsection{Summary of results}

Summarizing, we have in this section applied the current approach to discuss
evolutionary features of the standard cosmological parameters and calculated
their current values. Based on the estimated [3] acceleration commencement
look-back time $t_{ac}$, we found from a range of choices that $t_{ac}%
=4.07\pm0.05$ $Gyr$ gave estimates most consistent with observations for all
the parameters. Using this value of $t_{ac}$ we estimated the age of the
Universe $t_{0}$, the Hubble constant $H_{0}$. The results are consistent with
the latest observations.[38][6][39][40]. The model also generates two fields
$\Delta_{+-}$and $\Delta_{++}$ which we identified as the net pressure
(component) of dark energy and net matter/dark matter density component. We
have showed the two fields have the desirable characteristics. In particular
$\Delta_{+-}$ can accelerate the universe, while $\Delta_{++}$ can constitute
lumpy matter. The latter also provides dark matter halos with the correct
density profile that can produce the observed flat rotation curves [42] [43]
[44].. We have, futher, showed that the two fields $\Delta_{+-}$and
$\Delta_{++}$ give the observed respective relative percentages of dark energy
and of matter densities, in the universe [6]. Finally, in the process, we have
also introduced some new parameters including the cosmic period $\tau$, and
the synchronous time $t_{sy}$ and addressed the synchronicity problem. Table
2, below, summarizes our estimates of cosmological parameters, in the model.%

\begin{align*}
&
\begin{tabular}
[c]{|c|c|c|c|c|c|c|}\hline
$Input$ & \multicolumn{6}{|c|}{$Output$}\\\hline
$t_{ac}/Gyr$ & $t_{0}/Gyr$ & $H_{0}/kms^{-1}/Mpc$ & $\tau/Gyr$ &
$t_{sy-1}/Gyr$ & $\Omega_{de}$ & $\Omega_{m}$\\\hline
$\mathbf{4.07}_{-0.05}^{+0.05}$ & $\mathbf{13.76}_{-0.17}^{+0.17}$ &
$\mathbf{68.22}_{-0.85}^{+0.87}$ & $\mathbf{19.38}_{-0.24}^{+0.24}$ &
$\mathbf{13.86}_{-0.35}^{+0.17}$ & $69.7\%$ & $30.3\%$\\\hline
\end{tabular}
\\
Table2  &  :Summary\ of\text{ }model\prime
s\ calculated\ best\ estimates\ of\ cosmological\text{ }parameters.
\end{align*}

\section{Concluding remarks}

In this work we have presented a framework to discuss dynamics of a universe
with self-regulating features. The approach builds on the Friedman model by
introducing two ingredients. First, that basic-space is endowed with a
physical structure and this makes it one of the active drivers of the dynamics
of this universe, and second that matter-energy emerges as a perturbation to
the structure and to the dynamics of this universe. The underlying idea that
basic-space has structure leads to a flat geometry with a time-linear
expanding space as the ideally unperturbed dynamic equilibrium state. We have
described these contributions of basic-space to the dynamics of the modeled
universe. Later we added those contributions of free matter-energy, as
perturbations. The two ingredients are linked through a proposal referred to
as the Dynamic Equilibrium Protection Proposal (DEPP). According to DEPP
cosmic dynamics always seeks stability against perturbations of its dynamic
equilibrium state. We combined the two contributions into one framework to
describe the over-all dynamics. It describes a universe that oscillates
periodically about a time-linearly, monotonically expanding path. The model
isolates two density fields $\Delta_{++}$ and $\Delta_{+-}$ driving the dynamics

We have compared the features and characteristics of the modeled universe with
the observed physical Universe. We find the former reproduces very well the
general evolutionary profile of the observed Universe from the early big bang
phase through radiation and structure formation phases to the current
acceleration. The model is shown to have neither a horizon problem nor a
flatness problem. Its initial phase acceleration ends into a big bang and
matter creation. Similarly, the second acceleration, after structure
formation, begins predictably and will end predictably. Further, subject only
to observational constraints of when the current cosmic acceleration
initiated, the model is able to reproduce the standard current values of the
cosmological parameters, accurately, including the age, $t_{0}$, the Hubble
$H_{0}$ and the density parameters of dark energy $\Omega_{de}$ and that of
gravitating matter-energy fields $\Omega_{m}$ to a high accuracy consistent
with recent observations. The dark matter produced in the model has a density
profile that facilitates flat rotation curves in galaxy halos, consistent with observations.

We have demonstrated the model can address some standing puzzles. As an
example we have discussed the origin of the \textit{dimensionless
age}/s\textit{ynchronicity} problem and addressed it, in the process fixing
the various times in future when cosmic dynamics will satisfy $Ht=1$. With
regard to testability, the model predicts features that make it testable and
falsifiable. For example according to the model the beginning of cosmic
acceleration $t_{ac}\sim4Gyr$ ago, also marks both the minimum expansion rate
$\dot{a}_{\min}$ (see Figures 1 and 2) and the minimum of dark matter-energy
density content $\Delta_{++}$ (see Eqs. 6.6 and 6.7). Therefore the temporal
region of about $2-3Gyr$ centered at this point \ constitutes a phase referred
to here as the "Cosmic Valley", with special testable features. In the Cosmic
Valley: 1) Expansion rate tests such as Type 1A supernovae should register
lower than expected radial recession velocities of galaxies; 2) Galaxies
observed in this region should mostly be compact with little or no dark matter
halos. 3) The predicted primordial non-inflationary accelerated expansion of
the model may favor earlier than expected formation of structure. Finally, it
is expected that more tests, including CMB based, nucleosynthesis and
structure formation will improve the model.

\section*{Acknowledgments}
We would like to thank Fred Adams, Stephon Alexander, Niayesh Afshordi, Robert
Brandenberger, Gerald Dunne, Demos Kazanas, Ronald Mallett and Philip Mannheim
for useful comments and/or observations regarding this work.

\textit{This work was supported, in part, by the Swedish International
Development Cooperation Agency (SIDA) through the International Science
Program (ISP) grant N\b{o}\ RWA:01}

\section*{Appendix1}
\begin{figure}[h!]
\centering
\includegraphics[scale=0.4]{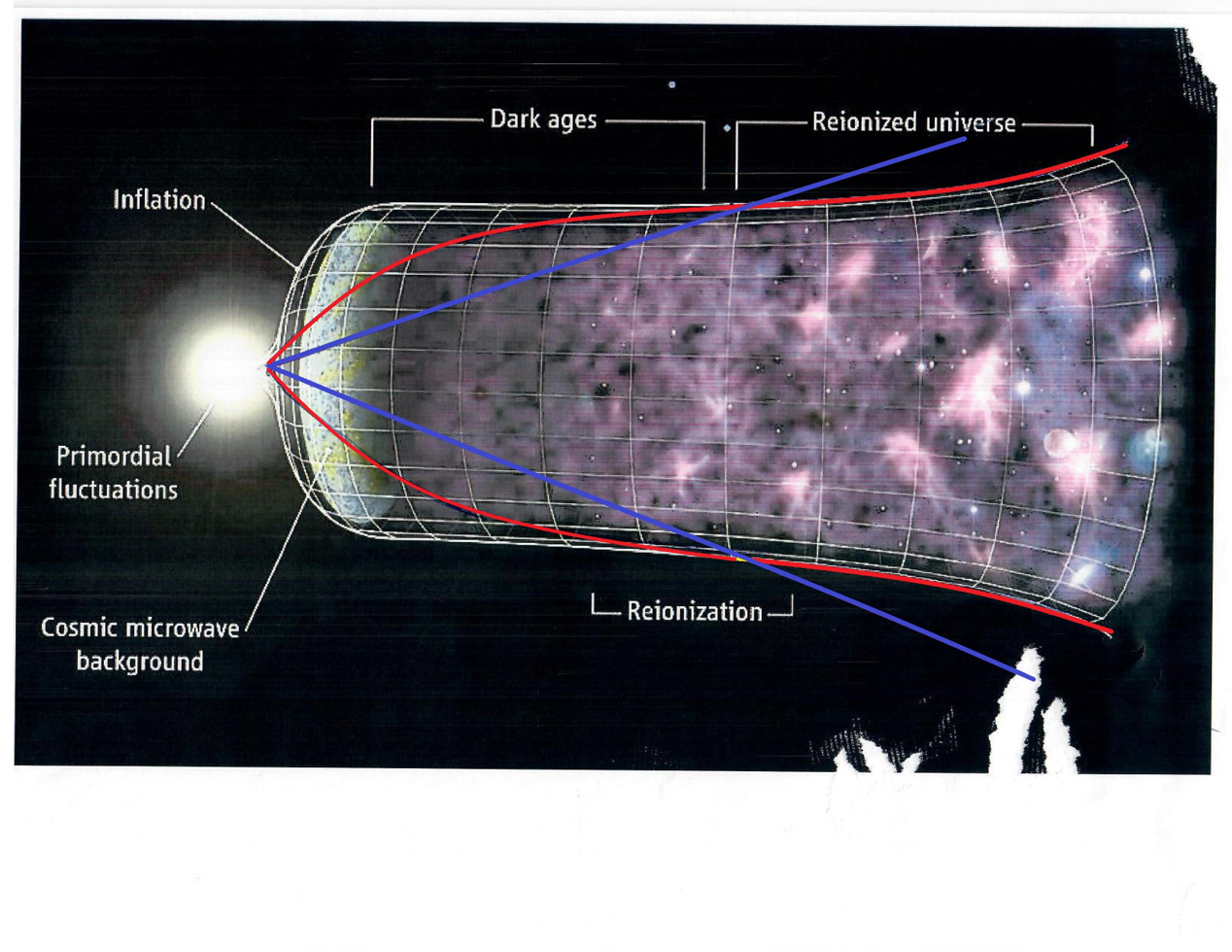}
\caption{A diagramatic sketch of cosmic evolution: based on lambdaCDM model (Courtesy of NASA/WMAP Science Team). Superimposed on it is our model (in red) snaking about a (blue not on scale) coasting path.}%
\label{Figure appendix.png}%
\end{figure}

\end{document}